\documentclass[12pt]{spieman}  
\usepackage{amsmath,amsfonts,amssymb}
\usepackage{setspace}
\usepackage{tocloft}
\usepackage{graphicx}

\usepackage{textcomp, gensymb}
\usepackage[detect-all, separate-uncertainty=true]{siunitx}
\usepackage{bm}
\usepackage[dvipsnames]{xcolor}
\usepackage{booktabs}

\def\brcurs{{\mbox{$\resizebox{.09in}{.08in}{\includegraphics[trim= 1em 0 14em 0,clip]{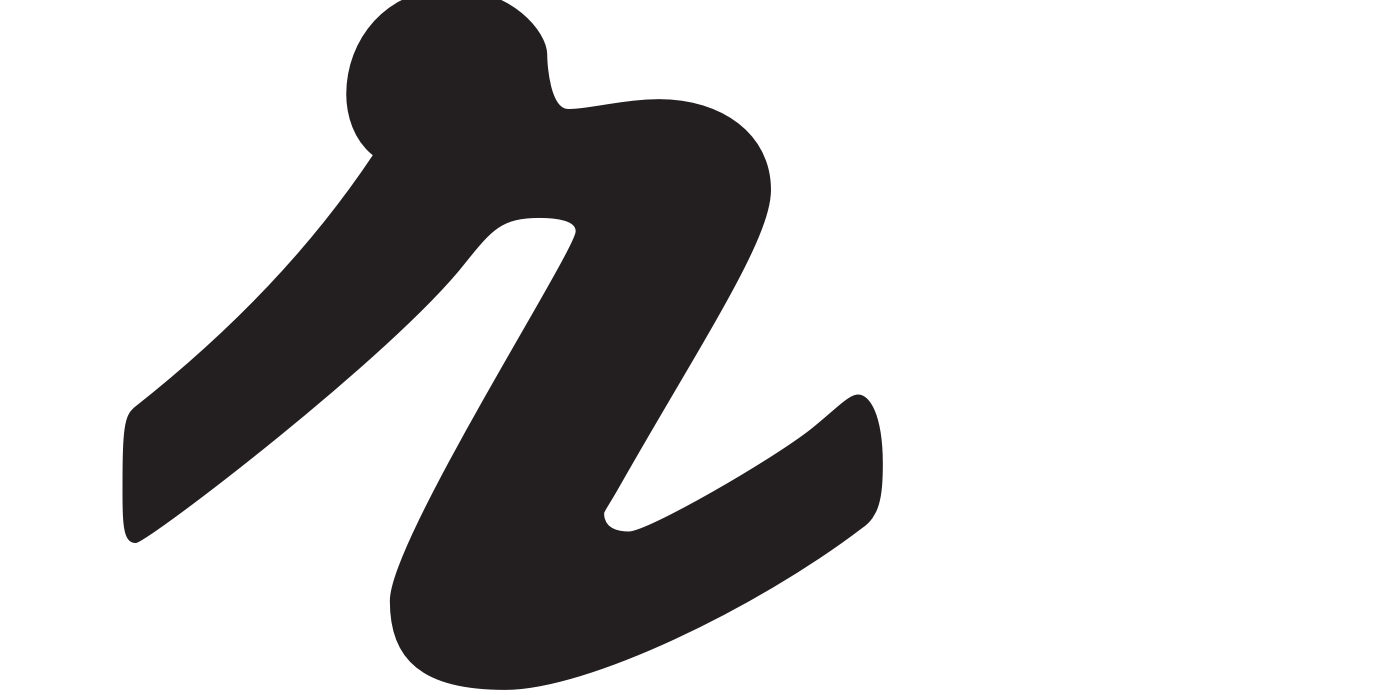}}$}}}

\title{Focal Plate Structure Alignment of the Dark Energy Spectroscopic Instrument}

\author[a,b]{Duan Yutong}
\author[c]{Joseph H. Silber}
\author[c]{Todd M. Claybaugh}
\author[a]{Steven P. Ahlen}
\author[d]{David Brooks}
\author[e]{Gregory Tarlé}
\affil[a]{Physics Department, Boston University, 590 Commonwealth Ave, Boston, MA 02215, USA}
\affil[b]{Physics Division, Lawrence Berkeley National Laboratory, 1 Cyclotron Rd, Berkeley, CA 94720, USA}
\affil[c]{Engineering Division, Lawrence Berkeley National Laboratory, 1 Cyclotron Rd, Berkeley, CA 94720, USA}
\affil[d]{Department of Physics Astronomy, University College London, Gower Street, London WC1E 6BT, UK}
\affil[e]{Physics Department, University of Michigan, 450 Church Street, Ann Arbor, MI 48109, USA}

\cftpagenumbersoff{figure}
\cftpagenumbersoff{table} 
\begin{document} 
\maketitle

\begin{abstract}

	The Dark Energy Spectroscopic Instrument (DESI) is under construction to measure the expansion history of the universe using the Baryon Acoustic Oscillation (BAO) technique. The spectra of 35 million galaxies and quasars spanning over $14000 \,\text{deg}^2$ will be measured during the life of the experiment. A new prime focus corrector for the KPNO Mayall telescope will deliver light to 5000 robotically positioned optic fibres. The fibres in turn feed ten broadband spectrographs. Proper alignment of the focal plate structure, mainly consisting of a focal plate ring (FPR) and ten focal plate petals (FPP), is crucial in ensuring minimal loss of light in the focal plane. A coordinate measurement machine (CMM) metrology-based approach to alignment requires comprehensive characterisation of critical dimensions of the petals and the ring, all of which were 100\% inspected. The metrology data not only served for quality assurance (QA), but also, with careful modelling of geometric transformations, informed the initial choice of integration accessories such as gauge blocks, pads, and shims. The integrated focal plate structure was inspected again on a CMM, and each petal was adjusted individually according to the updated focal plate metrology data until all datums were extremely close to nominal positions and optical throughput nearly reached the theoretically best possible value. This paper presents our metrology and alignment methodology and complete results for twelve official DESI petals. The as-aligned, total RMS optical throughput for 6168 positioner holes of twelve production petals was indirectly measured to be $99.88 \pm 0.12 \%$, well above the 99.5\% project requirement. The successful alignment fully demonstrated the wealth of data, reproducibility, and micron-level precision made available by our CMM metrology-based approach.
	
\end{abstract}


{\noindent \footnotesize\textbf{*}Duan Yutong,  \linkable{dyt@physics.bu.edu} }

\begin{spacing}{2}   

\section{Introduction}
\label{sec:intro}  
	
	\begin{figure}[tbh]
		\centering
		\includegraphics[scale=0.47]{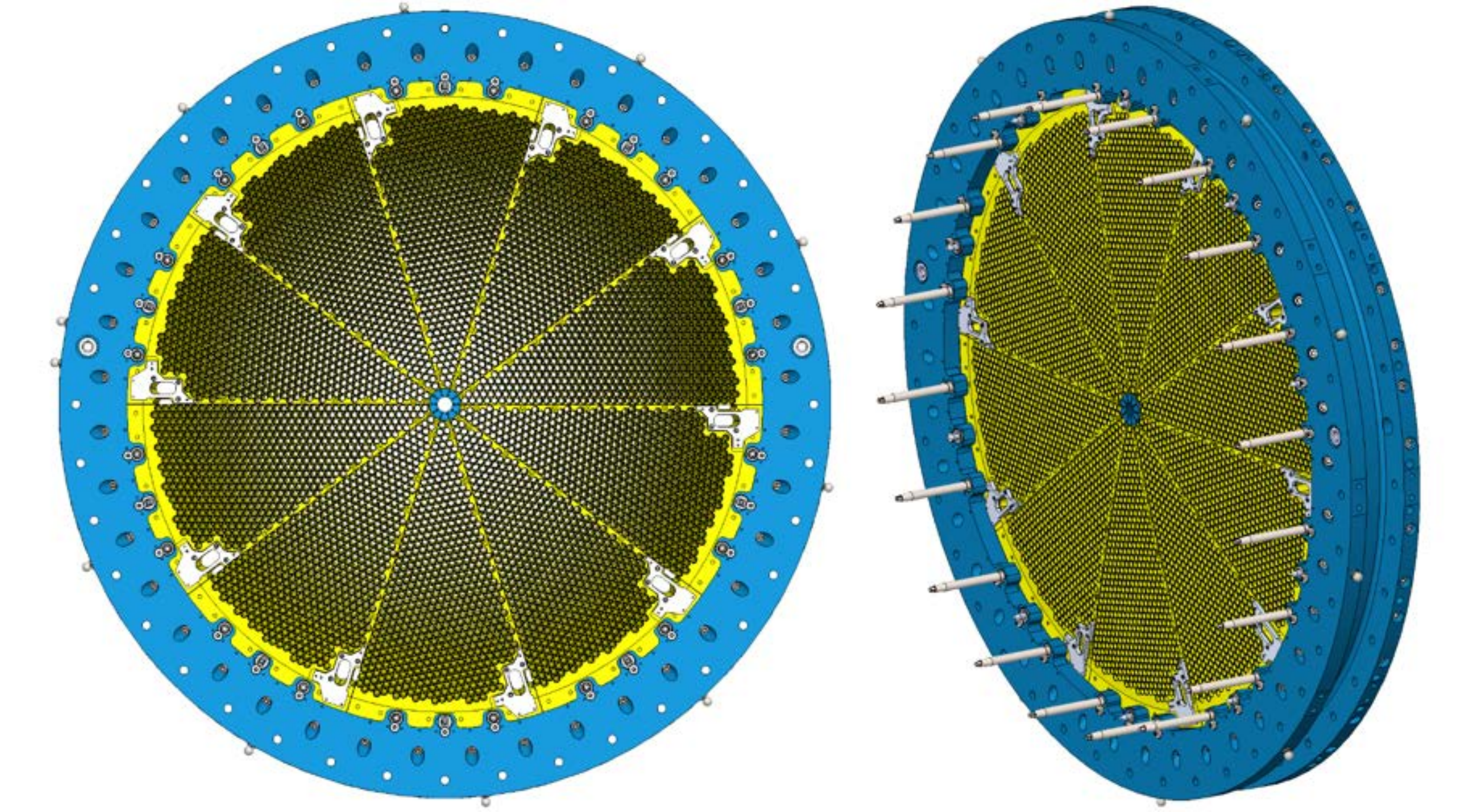}
		\caption{
		(Colour online) Design of the focal plate structure. Ten focal plate petals (FPP, rendered yellow) are mounted with guide pins and spikes and bolted onto the focal plate ring (FPR, rendered blue), joined at the centre by centre cap rings. Each petal hosts a guide focus alignment (GFA) sensor bolted to the GFA-FPP mount plate (rendered white) near the inner diameter of the ring. The focal plate structure is about \SI{1}{\meter} across in diameter.
		}
		\label{fig:fps_cad}
	\end{figure}
	
	The Dark Energy Spectroscopic Instrument (DESI) is the next-generation ground-based dark energy experiment to be installed on the 4-meter Mayall Telescope at the Kitt Peak National Observatory (KPNO) in Arizona. DESI will complement imaging surveys such as the Dark Energy Survey (DES) and the Large Synoptic Survey Telescope (LSST) to study the evolution of dark energy and growth of structures in the expansion history of the universe. Arrays of robotic positioners and fibre-fed spectrograph are capable of collecting spectra from 5000 targets simultaneously. During a five-year survey starting in 2020, DESI will make a three-dimensional map of over 35 million galaxies, covering $14000 \,\text{deg}^2$ of the sky and reaching to redshift $z \sim 3.5$. Using the Baryon Acoustic Oscillation (BAO) and Redshift-Space Distortion (RSD) techniques, DESI will provide new constraints on models of dark energy, modified gravity, inflation, and neutrino masses\cite{desi_fdr_part_1}.

	The focal plate is the main mechanical structure of the focal plane system, one of six subsystems of DESI\cite{desi_fdr_part_2, desi_overview_martini}.
	Given the optical design of the DESI corrector optics{\cite{desi_corrector}}, precise alignment between parts is critical to achieve the highest photons injection efficiency and minimise the loss of photons from targets.
	The focal plate structure must tightly constrain the optical fibres such that the fibre tips point in the direction of the chief rays, and that the patrol disks of the positioners maximally coincide with the aspherical focal surface. Every step in the process of building the focal plate was taken with these two goals in mind, from the petal design, stringent machining tolerances, to the accurate alignment of the focal plate structure.
	
\begin{figure}[tbh]
		\centering
		\includegraphics[width=\linewidth]{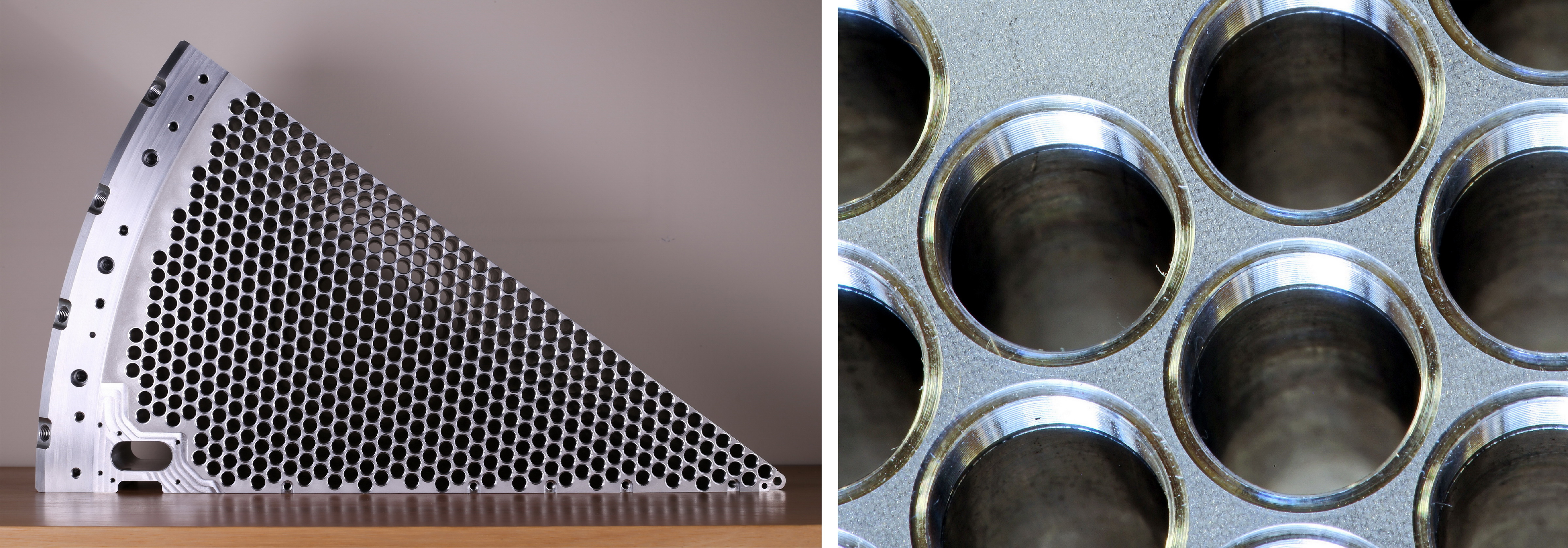}
		\caption{
			\label{fig:petal_photos}
			Focal plate petal machined from a block of aluminium before being black anodised. Left: petal front view. 514 positioner and FIF holes packed in a hexagonal pattern. The wedge is about half a meter long in radius. The GFA-FPP mount plate and the GFA are installed at the lower left empty slot. Right: petal hole details. Each petal hole consists of a precision bore of length \SI{20.5}{\milli\meter}, threading above the precision bore, and a circular precision spotface. The top rough surface is a result of sandblasting to reduce light reflection. Hole diameter is about \SI{8.31}{\milli\meter}. }
	\end{figure}

\begin{figure}[tbh]
		\centering
		\includegraphics[width=\linewidth]{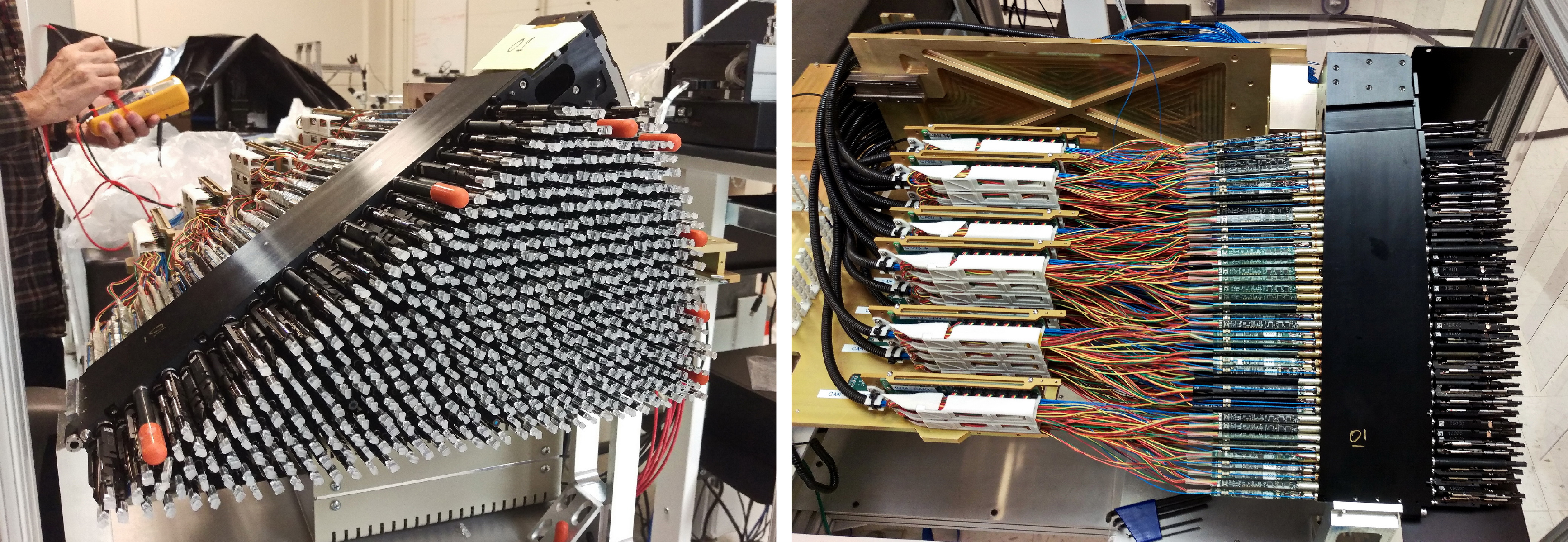}
		\caption{
			\label{fig:petal_integrated}
			(Colour online) Front view (left) and side view (right) of PTL01 fully populated with positioners and FIFs. Each positioner is constrained by the precision bore and oriented in a unique direction along the local chief ray, and its extension is determined by the precision spotface. Fibre tips, covered by white protective caps, extend out from the spotfaces by \SI{86.5}{\milli\meter} and lie on the focal surface. The positioners also have to built with impeccable straightness and length to ensure correct fibre tip positions.}
	\end{figure}

	Fig. \ref{fig:fps_cad} shows a computer-aided design (CAD) model of the DESI focal plate structure, mainly consisting of the focal plate ring (FPR) and ten focal plate petals (FPP) \cite{lambert_fp_integration}. 5000 robotically actuated positioners and 100 similarly sized field illuminated fiducials (FIF) are arranged in a hexagonal pattern with a $\SI{10.4}{\milli\meter}$ pitch, evenly distributed in ten pie-slice-shaped, identical petals. Each science fibre can be positioned freely within the \SI{12}{\milli\meter} patrol disk of a positioner. FIFs in the focal plane provide point light sources as references and help to reduce optical distortion and improve fibre positioning accuracy when viewed by the fibre view camera (FVC)\cite{desi_fdr_part_2} located in the Cassegrain cage. Also evenly distributed in a circular pattern are ten guide focus alignment (GFA) sensors, which measure the telescope pointing and focus as well as the tip/tilt of the focal surface. As shown in Fig. \ref{fig:petal_photos}, each petal has 514 holes in which the positioners and FIFs are installed. The precision bore in each hole is reamed at a unique angle along the local chief-ray direction to constrain the orientation of the fibre positioner. The precision spotface atop each hole perpendicular to the cylinder axis determines how far in the positioner can be screwed, and places the fibre tip into the depth of focus of the corrector optics (Fig. \ref{fig:petal_integrated}). The petals and the ring were anodised black, and the top spherical dome of the petals were sandblasted in order to reduce reflection of stray photons from undesired targets. The centre cap rings join all ten petals at their noses and enhance the rigidity of the focal plate structure.
	
	We performed 100\% inspection of the petals and the ring with coordinate measurement machines (CMM), which yielded a complete characterisation of all critical dimensions prior to alignment. It became a natural choice to capitalise on the existing inspection data and adopt a metrology-based approach for our alignment goals. The focal plate design allows for individually aligning each petal by choosing points of contact from three existing slots and by varying the sizes of gauge blocks and petal-ring pads. The GFA sensor fixtures, i.e. GFA-FPP mount plates (Fig. \ref{fig:gfa_alignment_shims}), also must be aligned to ensure correct position and orientation of the GFA sensors. Our alignment approach was first validated in August 2017 by test-fitting five production petals in the ring and arriving at nearly optimal configurations. All 12 production petals were aligned in two runs in November 2017 and January 2018, and 11 of them had GFA-FPP mount plates aligned.
	
	Having successfully completed the focal plate alignment by January 2018, we present our methodology and results in detail in this paper. \autoref{section:parts_inspection} provides the methods used for FPP and FPR inspection. \autoref{section:throughput_eval} introduces optical throughput modelling and evaluation, and various alignments found by our custom optimiser in which throughput was evaluated. In \autoref{section:integration_alignment}, procedures for aligning petals and the GFA-FPP mount plates are described. Complete metrology and alignment results can be found in \autoref{section:results_discussions} along with discussions. Finally \autoref{section:conclusions} summarises our findings.
	
\section{Parts Inspection}
\label{section:parts_inspection}
	
	\subsection{FPP Inspection}
	
		Metrology data of the petals, in particular the holes, are direct indicators of future science performance, and were extensively analyzed and utilised for focal plate alignment. All DESI prototype and production petals were fabricated at the Scientific Instrument Facility (SIF) of Boston University (BU) with a 5-axis machine, and coated with organic black anodising by Plating for Electronics Waltham, MA. The in-house inspection of petals started with gluing three datum tooling balls made of silicon nitride to the petal with epoxy widely apart, which established a fixed frame of reference for the as-built petal. Bearing cartridges (a part of the fibre positioners which screws into petal holes) were screwed in by hand to check the hole threads and fit. The precision bore diameter was measured with a Mitutoyo 3-point holtest near the middle section of the cylinder to $\SI{\pm 1}{\micro\meter}$ accuracy, before and after anodising to track the actual anodising thickness and decrease in diameter. The nutation angle (polar angle in spherical coordinates system of the CAD model) was measured on a Mitutoyo optical comparator to about $\SI{\pm 0.02}{\degree}$ with a pin gauge of appropriate diameter tightly inserted into the precision bore and parallel to the cylinder axis. The $z$-coordinate of the precision spotface centre was measured with a Mitutoyo height gauge. A precision sphere of appropriate diameter precisely sits on the inner ledge of the spotface, the height gauge touches down on the sphere from above reading the highest point, and the $z$-coordinate of the spotface centre could be derived with a bit of simple geometry calculation. A Brown \& Sharpe Quadra-Chek CMM was also employed to measure the flatness of plane datums on petals until the machine went out of order. The first four petals were hand-checked for 100\% of the holes, and later ones from 10\% to 50\%. 
	
		BU developed the CMM automation programme in collaboration with Zeiss Industrial Metrology Marlborough, MA, and the petals were 100\% inspected by a Contura CMM running Calypso software. The most challenging features in petal metrology are the precision bores and spotfaces; the latter are minuscule in size and necessitate the smallest ruby styli available (\SI{0.3}{\milli\meter} in diameter). Each precision bore was measured by the probe scanning along two circular paths near the top and the bottom of the bore. Three iterations of the $3\sigma$ clipping filter were applied to the cylinder fit to make it resistant to outliers. The cylinder fit yields the diameter and axial direction of the bore. The programme looked for the critical spotface relative to the corresponding cylinder already found beneath it, and took a circular path on the flat surface which defines the plane. The relative definition of the spotface with respect to already measured bores guarantees the probe almost always completes a perfect circular path without running off the plane into the hole or colliding into the sandblasted surface above the spotface. The spotface centre was defined as the intersection between the cylinder axis and the spotface plane. A single run of the CMM programme takes less than 20 hours. Besides the 514 holes, other commonplace features were scanned in the usual manner. Calypso expresses the direction of cylinder axis $\hat{\bm{n}}$ as two projection angles onto the $XZ-$ and the $YZ-$planes, $A_1 = \tan ^{-1} \frac{x}{z}$ and $A_2= \tan ^{-1} \frac{y}{z}$, where $x, y, z$ are the Cartesian components of the axial unit vector $\hat{\bm{n}}$, namely $x \equiv \hat{\bm{n}} \cdot \hat{\bm{i}}, y \equiv \hat{\bm{n}} \cdot \hat{\bm{j}}, z \equiv \hat{\bm{n}} \cdot \hat{\bm{k}}$, and $\hat{\bm{n}}$ is the measured, actual axial direction. The specification of hole angles is by the nutation and precession angles $(\theta, \varphi)$, which are the ordinary spherical polar and azimuthal angles. The fitted cylinder angles $(A_1, A_2)$ were converted to nutation and precession angles $(\theta, \varphi)$ as
		\begin{align}
			\theta  &= \tan^{-1} \dfrac{\sqrt{x^2+y^2}}{z} = \tan^{-1}\sqrt{\tan^2 A_1 + \tan^2 A_2}\\
			\varphi &= \tan^{-1} \dfrac{y}{x} = \tan^{-1}\dfrac{\tan A_2}{\tan A_1} \, .
		\end{align}
		
		In CMM metrology, the usual method of establishing a coordinate system, or ``alignment" in CMM terminology, is by providing the three datums A, B, and C in the drawing to the CMM software. However, for a part as complicated as the petal where over 500 holes are closely packed all carrying fine features, it was found during CMM programme development that a naïve ``ABC alignment" resulted in half of features being out of tolerance due to imperfect datums. Specifically, datum A of the petal is flat to \SI{10}{\micro\meter}, while datums B and C are only specified to \SI{100}{\micro\meter} as they do not mate with any other surface when integrated and are therefore not critical. Considering the fact that by design each petal was to be individually adjusted to achieve the desired alignment, instead of the usual ABC alignment, we found a more reasonable ``best-fit" coordinate system in which the metrology data became more meaningful and informative. This best-fit coordinate system (denoted ZBF alignment) was generated in Calypso by performing a geometric best-fit of the measured data to 514 cylinders and 514 spotfaces simultaneously, and was checked outside Calypso to recover the transformation parameters in a standard convention and verify the transformed metrology results. 
	
	\subsection{FPR Inspection}
	
		BU contracted with Dial Machine Rockford, IL to manufacture the FPR and apply 2-step inorganic black anodising. Eleven datum tooling balls were glued to the outer rim of the ring as fixed references. Although the ring also requires high precision machining relative to its size (\SI{10}{\micro\meter} flatness across \SI{1}{\meter} diameter), it has a much simpler design and its features are much easier for CMM probes to work with because of relatively large sizes. BU verified the flatness of the top surface of the ring using a straight steel beam and a granite table,  to \SI{40}{\micro\meter} before shipping it to LBNL for CMM inspection and match-drilling with the focal plate adapter. 
		
		LBNL inspected the FPR on a Zeiss Accura CMM. Important features inspected included the flatness of the top surface (datum A), cylindricity of the ring (datum B), position and diameter of two counterbores (datum C), and perpendicularity between datums A and B. The 11 datum tooling balls and 10 slotted bushings were also measured. The coordinate system was established by supplying the three ABC datums. The aluminium ring was laid directly on the CMM granite table without any supporting jack stands because of observed gravitational sagging, and this set-up continued throughout alignment activities. During on-mountain installation, the FPR will be bolted to the DESI corrector barrel, which is made of aluminium and steel, flat to \SI{150}{\micro\meter}, and will provide adequate support similar to that from the granite table. Therefore, not using jack stands was a superior way to reproduce in the inspection lab the support provided by the actual instrument structure for the ring.
		
\section{Science Performance Evaluation}
\label{section:throughput_eval}
	
	The quality of each petal was to be assessed in terms of not only the specifications, but more importantly, the science performance. The science performance must be determined before the parts could be accepted and used further along the integration process. Optical throughput is a metric which quantifies how much light from the target enters the fibre and how well the petals will perform in observation, but it was extremely prohibitive to measure it directly due to many constraints. It would take months to install the fibre positioners needed to perform direct measurement; fibre positioner availability was very limited; plus, it would pose a major risk to the entire project if a large number of production positioners were repeatedly installed and uninstalled just for this test.
	
	Although the only basis for projecting the optical throughput was  inspection data, it would suffice for the purpose of assessing the petal quality. By comparing the actual geometry to the nominal, all fibre orientations, fibre tip positions, and thus the theoretical throughput of the petal could be calculated from the optical and geometrical model of the focal plate. Here optical throughput is defined as the percentage of incident photons along the local chief ray which successfully enter the fibre. Features directly impacting throughput were the diameters and axial directions of precision bores, and the positions of spotface centres. This indirect measurement relies on a simple optical model and only required a few reasonable assumptions.
	
	Our optical model considers the loss of incident light incurred by two factors, based on how DESI systems engineering breaks down the error budgets \cite{desi_sys_eng}. One factor is angular tilt, the combined angular deviation of the orientation of the fibre from the local chief ray. The other is defocus, the absolute distance deviation of the fibre tip from the aspherical focal surface along the chief ray direction.
	Each fibre positioner has a patrol disk of \SI{12}{\milli\meter} radius, in which the fibre can be shifted while maintaining a constant orientation. As far as evaluating throughput is concerned, the fibre is always assumed to be at the centre of the patrol disk; lateral movements within the patrol disk certainly do introduce additional tilt and defocus, but they are taken into account in the system error budgets in a separate analysis.
	Since the error of the fibre tip position is orders of magnitude smaller than the patrol radius, in which the fibre orientation stays constant by design, it is safe to assume the local chief ray has a constant direction when the fibre tip deviates from its designed point. In addition, we neglect any manufacturing imperfections of the fibre positioner and assume the orientation of the fibre is perfectly parallel to the cylinder axis of the precision bore, which constrains the fibre-carrying positioner. In other words, the fibre tilt is equal to the precision bore tilt. 
	
	Recall that the nutation and precession angles $(\theta, \varphi)$ can be converted to unit vectors in Cartesian coordinates as $\hat{\bm{n}} (\theta, \varphi) = \sin\theta\cos\varphi \hat{\bm{i}} + \sin\theta\sin\varphi \hat{\bm{j}} + \cos\theta \hat{\bm{k}}$. We may calculate the actual cylinder axis of the precision bore $\hat{\bm{n}} \equiv \hat{\bm{n}} (\theta, \varphi)$ and the nominal $ \hat{\bm{n}}_0 \equiv \hat{\bm{n}} (\theta_0, \varphi_0)$, where $(\theta, \varphi)$ are the measured angles and $(\theta_0, \varphi_0)$ are the specified angles. The tilt is then defined as
	\begin{equation}
		\delta \hat{\bm{n}} (\theta, \varphi) \equiv \cos^{-1} ( \hat{\bm{n}} \cdot \hat{\bm{n}}_0 ) = \cos^{-1} \left[ \hat{\bm{n}} (\theta, \varphi) \cdot \hat{\bm{n}} (\theta_0, \varphi_0) \right] \, . 
	\end{equation}
	From a DESI fibre focal ratio degradation (FRD) study by Jelinsky\cite{frd_tilt}, where optical throughput was measured as the fraction of full-cone encircled energy at \SI{625}{\nano\meter}, we extracted the dependence of optical throughput on tilt as a quadratic polynomial,
	\begin{equation}
		T_\text{tlt} (\delta\hat{\bm{n}}) = -0.0133 \cdot \delta\hat{\bm{n}}^2 - 0.0175 \cdot \delta\hat{\bm{n}} + 1
	\end{equation}
	where the tilt $\delta\hat{\bm{n}}$ is in degrees. $T_\text{tlt} (\delta\hat{\bm{n}})$ gives negative throughput for large tilt, and is limited at 0 as a lower bound. To calculate defocus, let $\bm{r}$ and $\bm{r}_0$ be the actual and nominal spotface centres. The positioner design places the fibre tip exactly \SI{86.5}{\milli\meter} above the spotface centres along the cylinder axis. Assuming positioners are perfect in length and orientation, the actual and nominal fibre tip positions are
	\begin{align}
		& \brcurs = \bm{r} +  86.5 \cdot \hat{\bm{n}} \\
		& \brcurs_0 = \bm{r}_0 + 86.5 \cdot \hat{\bm{n}}_0
	\end{align}
	all in units of \SI{}{\milli\meter}. Defocus is the absolute distance deviation of fibre tip along the chief ray direction,
	\begin{equation}
		\delta f (\theta, \phi, \bm{r}) = \left|  \delta \brcurs \cdot \hat{\bm{n}}_0 \right| = \left| (\brcurs - \brcurs_0) \cdot \hat{\bm{n}}_0 \right| \, .
	\end{equation}
	From DESI fibre defocus data by Besuner and Sholl \cite{desi_sys_eng}, the throughput as a function of defocus can be modelled as a 5th-order polynomial,
	\begin{eqnarray}
		T_\text{dfc} (\delta f) = 
					- 1.804 \cdot 10^{-14} \cdot \delta f^5
					+ 1.593 \cdot 10^{-11} \cdot \delta f^4
					- 5.955 \cdot 10^{-10} \cdot \delta f^3 
					- 3.433 \cdot 10^{-6}  \cdot \delta f^{2} \nonumber\\
					+ 3.251 \cdot 10^{-7}  \cdot \delta f 
					+ 1
	\end{eqnarray}
	where $\delta f$ is in units of \SI{}{\micro\meter} and always measured as an absolute value of the distance deviation. $T_\text{dfc} (\delta f)$ also gives negative throughput for large defocus, and is bounded at 0 from below. Finally, the total optical throughput for a given fibre as a function of the actual nutation, precession, and spotface centre is the product
	\begin{equation}
		T(\theta, \varphi, \bm{r}) =  T_\text{tlt} (\delta \hat{\bm{n}}) T_\text{dfc} (\delta f)
	\end{equation}
	which is a function of bore orientation and spotface centre position only.
	
	Inspection data were analysed in the geometric best-fit (ZBF) alignment, which rendered almost all features measured within their specifications and was clearly superior to the ABC alignment, but it was unknown to what extent the best-fit alignment would optimise optical throughput, or if there exists a more preferable alignment in which throughput would be higher. To this end, a custom alignment fitting programme was independently written in Python 3.6 using the SciPy.optimise module and the Sequential Least SQuares Programming (SLSQP) algorithm\cite{scipy} to optimise solely the optical throughput according to our optical model and find a throughput-optimising alignment (denoted TPT). Also implemented was a spotface-fit alignment (SPT), another geometric best-fit alignment which fits only to the 514 spotface centre and in principle ought to be inferior to the best-fit alignment (ZBF). The optimiser finds the ideal transformation parameters relative to the ABC alignment by varying three translational and three rotational degrees of freedom and minimising the total mean throughput loss of 514 fibres. The petals are stiff enough to be considered as rigid bodies, and the transformation formalism used is
	\begin{equation}
		\bm{r}^\prime = R \bm{r} + \bm{T} = R_z(\gamma)R_y(\beta)R_x(\alpha) \begin{pmatrix}
			x\\
			y\\
			z
		\end{pmatrix}
		+
		\begin{pmatrix}
			T_x\\
			T_y\\
			T_z
		\end{pmatrix}
	\end{equation}
	where $R_x, R_y, R_z$ are $3\times3$ Cartesian rotation matrices in the usual convention\cite{arfken2011mathematical}. In addition to transforming the metrology data to other coordinate systems using user-defined criteria, this fitter was also used to recover the transformation parameters of the best-fit (ZBF) alignment and verify the transformed metrology results. The petal metrology data set a theoretical upper bound on the optical throughput of the petal, as all 6 degrees of freedom are free parameters whereas in reality, possible petal adjustments will not span the entire parameter space. The Python code for the analyses is publicly available online\footnote{\href{https://github.com/duanyutong/desifp}{https://github.com/duanyutong/desifp}.}.

\section{Integration and Alignment Procedures}
	\label{section:integration_alignment}
	
	\subsection{Integration Tests}
	\label{subsection:integration_alignment}
		
		Integration tests took place with the FPR and the first three production petals in June 2017, and again with the first five petals in August 2017 before official alignment activities, all at LBNL in Berkeley, CA. During these tests, we examined the deformation of the FPR under uneven load and under different torques of the petal fasteners, investigated the reproducibility of alignment given our integration procedures, validated our software tools, and took special note of whether or not bolting the centre cap rings as the final step altered the alignment. We also checked how different mounting locations changes petal alignment, but since petals would be installed at fixed locations and aligned individually for the particular location on a petal-by-petal basis, this was not a major concern.
		
		A baseline measurement of the FPR alone was performed before installing any petal. Petals were installed without use of any guide pins or spikes, the fasteners hand-tightened followed by torquing to \SI{19}{\newton\meter} and \SI{28.5}{\newton\meter}. FPR flatness, concentricity and shifts of the datum tooling balls were measured after one, three, and five petals were installed. Petals were as evenly spaced as possible across ten petal mounting locations. A new coordinate system was generated from FPR datums for each measurement. These tests helped us tremendously to revise our alignment procedure and finalise the alignment adjustment scheme, which involved new torque specifications, the addition of petal datum A shims, and a redesigned nose shim structure which retains the centre cap rings.
		
	\subsection{Petal Alignment}
	
		\begin{figure}[tbh]
			\centering
			\includegraphics[width=\linewidth]{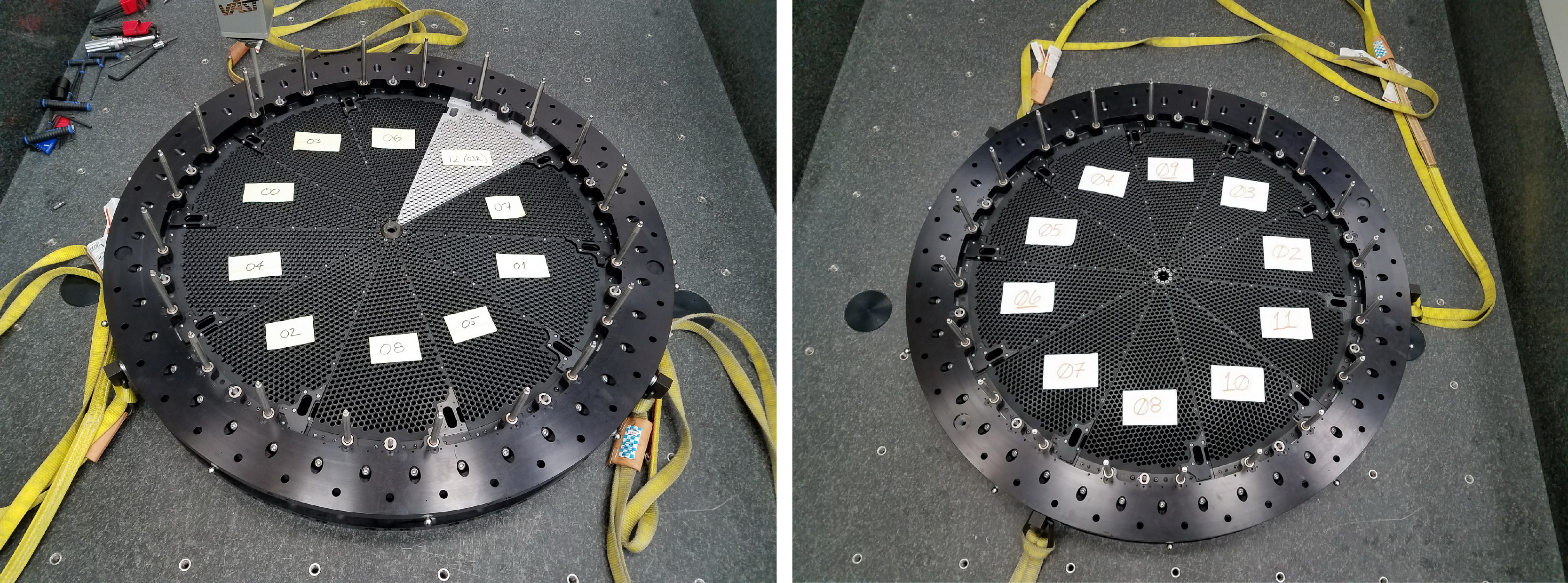}
			\caption{
				\label{fig:alignment_10_petals}
				Alignment of 12 official petals were done in two batches. Left: the first batch in November 2017 included 9 official petals, PTL00-08, and a rejected spare. Positioner integration began with PTL00 and 01 immediately afterwards. Right: the second batch in January 2018 included 10 official petals, PTL02 to 11. In both runs, guide spikes and pins were used, and alignment procedures were consistent.}
		\end{figure}
		
		Alignment of 12 production petals was done in two batches due to the project schedule. The first batch in November 2017 included the first nine petals, PTL00 through 08, after which petals PTL00 and 01 were handed to the LBNL technician team for positioner integration. A spare petal, supplied by BU and rejected in the manufacturing process due to imperfect quality, filled the last empty mounting location to maintain the proper weight distribution across the ring, making the ring fully occupied. The second batch included 10 production petals, PTL02 through 11. The alignment procedures were consistent for these two batches, and the 8 petals which were present both times saw improvements at the end of the second run. Aligned petals and ring are pictured in Fig. \ref{fig:alignment_10_petals}.

\begin{figure}[tbh]
			\centering
			\includegraphics[width=\linewidth]{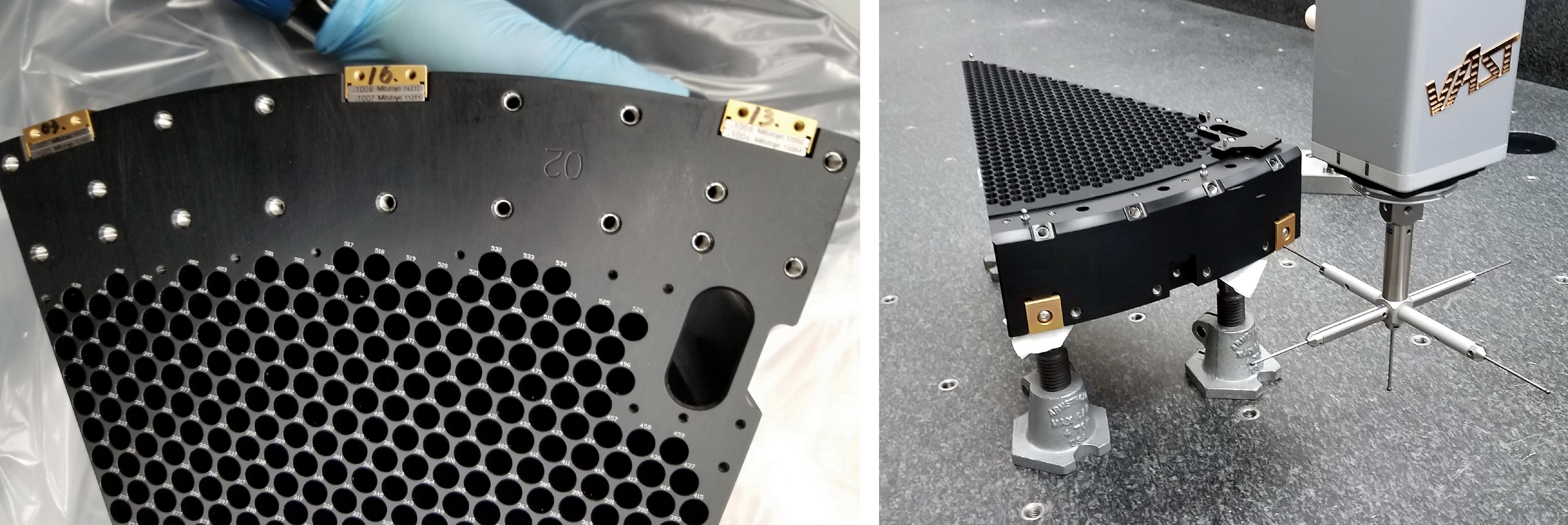}
			\caption{
				\label{fig:petal_ring_pads}
				Left: Three slots are available for petal-ring mount pad installation on a petal. By loosening the pad, the gauge blocks it houses can be slid out and hot swapped. Each pad is hand-measured and numbered to match the depth of each slot. Right: The three-slot scheme was found to be an over-design. Only the left and right slots were used, the middle left empty for aligning all petals. The petal-ring mount pad and GFA-FPP mount plate were surveyed on a CMM before petal alignment in the FPR.}
		\end{figure}

		Only two out of three petal-ring mounting pad slots, the left and the right, were used; the centre slot was left empty for all petals to avoid over-constraining the assembly and add clarity to our control over points of physical contact (Fig. \ref{fig:petal_ring_pads}). Two of the four top mounting screws were dropped as well for the same reason, leaving only the left and the right ones along with four bottom mounting screws to hold the petal to the ring. As a starting point, mounting accessories, such as petal-ring mounting pads, gauge blocks, and shims, were pre-selected based on pre-alignment metrology data by calculating their nominal sizes to achieve the desired alignment. Petals were installed with guide pins and spikes and torqued to \SI{4.0}{\newton\meter} for the horizontal bolts and \SI{19.8}{\newton\meter} for the $45\degree$ skewed bolts to keep rotational torque on petals at zero. The three datum tooling balls on every petal were measured in the shared focal plate coordinate system, generated from FPR datums. From the positions of three datums, the actual position, orientation, and throughput of the petal were evaluated. Based on the petal-ring geometry, the custom alignment optimiser inferred possible outcomes of varying the gauge blocks, recommended the next step adjustment which would maximise throughput, and showed the hypothetical improvement in throughput.

		Running independently in parallel with the Python script was a suite of spreadsheets, which served as a more user-friendly and transparent pipeline for CMM data processing and incorporated an alternative, approximated calculation as a cross-check. Petals were dismounted and adjustments were made accordingly by swapping out the gauge blocks and shims. Then a new coordinate system was established and the datums measured again. This iterative process of adjustment and verification was repeated several times for all petals, and the alignment progress was tracked closely until all datums were extremely close to nominal positions and the optical throughput nearly reached the theoretical maximum, set only by the petal machining quality. For a given petal, the RMS throughput of all holes is $\widetilde{\eta} = (\frac{1}{514} \sum_{i=1}^{514} \eta_i^2 )^{1/2}$. The alignment which produced the best RMS throughput was considered final and to be always reproduced in the future. 
		
		At last, the centre cap rings were installed with appropriate nose shims, and the focal plate surveyed again to verify that the alignment remained unchanged. The nose shims were carefully adjusted to avoid any distortion by the centre cap rings, and this process repeated until no distortion was present.
	
	\subsection{Guide Focus Array Alignment}
	
\begin{figure}[b]
			\centering
			\includegraphics[width=\linewidth]{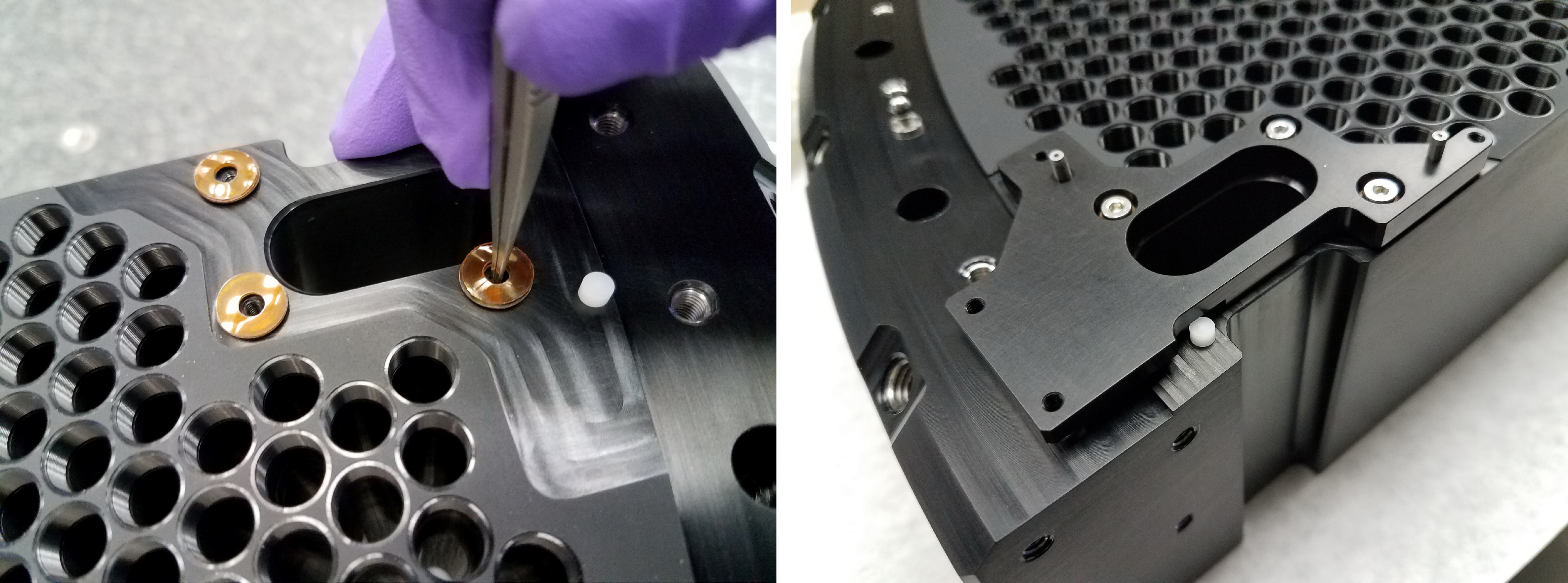}
			\caption{
				\label{fig:gfa_alignment_shims}
				The GFA-FPP mount plate is biased against shoulder pins and bolted down through three bores. Plastic shims were placed around the bolts beneath the bores to adjust the plate. The goal was to align the top surface of the plate such that it coincides with the nominal plane.}
		\end{figure}

		The GFA-FPP mount plate is an adjustable fixture between the petal and the GFA sensor. There are three bores on the the GFA-FPP mount plate; under each a stack of plastic shims can be added as pictured in Fig. \ref{fig:gfa_alignment_shims}. The bore centres, defined as the intersections between bore axes and the top face of the plate, were taken as reference datums. The GFA-FPP plate was aligned in the best-fit (ZBF) alignment relative to the three datum tooling balls glued to the petal, such that if a petal was aligned in the ring properly, the GFA-FPP plate would automatically follow into the correct position without requiring another round of alignment.
		
		For GFA-FPP mount plate alignment, the most reliable method is of course ordinary least squares (OLS), i.e. minimising the residual sum of squares of 3 datums $RSS = \sum_{i=1}^3 (\bm{x}^i - \bm{x}_0^i)^2 $ with respect to their nominal positions. However, it is not always possible to freely align the GFA-FPP mount plate to a perfect position given the mechanical design. For practicality and efficiency, we identified what was critical for the science task and opted another easier-to-apply strategy. As long as the GFA sensors are placed at the focal surface by the GFA-FPP plate, any small translation or rotation within the sensor plane does not significantly impact the science performance of the GFA. This means that the GFA-FPP plate needs to be aligned such that the top surface lies in the nominal plane as designed in the CAD model, and we only need to minimise the normal component of residual squares. Per engineering convention, the RMS of squared normal deviations was chosen as the loss function to be minimised instead of the sum, both mathematically equivalent. This minimisation in itself should result in a close match between actual and nominal positions of the GFA-FPP plate.
		
		The normal component of squared residuals was calculated in two ways. Let the nominal positions of three datums be $(\bm{x}_0^1, \bm{x}_0^2, \bm{x}_0^3)$, where the subscript 0 indicates nominal and the upper index indicates datum number, and the actual positions be $(\bm{x}^1, \bm{x}^2, \bm{x}^3)$. Each $\bm{x}^i$ is a 3-component vector. The normal vector of the GFA-FPP plate can be found by taking the cross product $\bm{n} \equiv (\bm{x}_0^1 - \bm{x}_0^2) \times  (\bm{x}_0^1 - \bm{x}_0^3)$, and the unit normal vector $\hat{\bm{n}} = \frac{\bm{n}}{\| \bm{n} \|}$. Then the normal deviation of each datum is given by the projection of datum position deviation along the normal vector, 
		\begin{equation}
			\delta d^i_\perp = (\bm{x}^i - \bm{x}_0^i) \cdot \hat{\bm{n}} \, .
		\end{equation}
		and the RMS squared normal deviation of three datums is
		\begin{equation}
			\widetilde{\delta d}_\perp = \sqrt{\frac{1}{3} \sum_{i=1}^3 (\delta d^i)^2} \,
		\end{equation}
		for a given GFA-FPP plate. This calculation involves coordinate transformation between the local coordinate system of the petal and that of the focal plate set by the FPR, and was done in Python in the same manner the petal alignment routine was written.
		
		Again, a spreadsheet was in place to perform a different calculation independently, and was used primarily during alignment for convenience. If $xy$ rotation with respect to the $z$-axis is neglected, we may consider only the radial $r$ and $z$ coordinates of each datum. Let the nominal and actual radial coordinates for each datum be $r_0^i = \sqrt{x_0^{i2} + y_0^{i2}}$, $r^i = \sqrt{x^{i2} + y^{i2}}$, and similarly we have $z_0^i$ and $z$. Let the nominal inclination angle of the GFA-FPP top face be $i_0$. Assuming the actual and nominal positions are in the same cross-section plane of the incline, we may write the normal deviation as
		\begin{equation}
			\delta d^{i \prime}_\perp 
				= \delta r^i \sin i_0 + \delta z^i \cos i_0 
				= (r^i - r_0^i) \sin i_0 + (z^i - z_0^i) \cos i_0 \, .
		\end{equation}
		Then the RMS squared normal deviation would be calculated in the same manner as $\widetilde{\delta d^\prime}_\perp = [ \frac{1}{3} \sum_{i=1}^3 (\delta d^{i \prime})^2 ]^{1/2}$.
		
		Minimising the squared normal deviations should at the same time minimise the inclination angle errors, and we measured and compared the inclination angle of top face of the plate to the nominal. Additional sanity checks included the true vector residual sum of squares $RSS = \sum_{i=1}^{3} (\delta\bm{x}^i)^2$ in \SI{}{\milli\meter\squared},  and the root mean of $RSS$
		\begin{equation}
			\widetilde{\delta d}_{rss}
				= \sqrt{\frac{1}{3} RSS}
				= \sqrt{\frac{1}{3} \sum_{i=1}^{3} (\delta\bm{x}^i)^2}
		\end{equation}
		in \SI{}{\milli\meter}, which is simply equal to the RMS vector deviations and comparable to the previous $\widetilde{\delta d}_\perp$ and $\widetilde{\delta d^\prime}_\perp$ calculated from normal projections.

\section{Results and Discussions}
\label{section:results_discussions}
	
	\begin{table}[t]
		\centering
		\caption{
			\label{table:petal_metrology}
			Petal metrology statistics for 12 production petals in the best-fit alignment. Columns are mean deviations ($\overline{\delta}$) from nominal diameter, spotface centre $x$, $y$, $z$, nutation, and precession for 514 holes of each petal. Parentheses in the header enclose plus-minus tolerances or science requirements. The symmetric errors in the values are $1\sigma$ standard deviations.}
		\begin{tabular}{
				c
				S[table-format=-2.1(2)]
				S[table-format=-2.1(3)]
				S[table-format=-2.1(3)]
				S[table-format=-2.1(3)]
				S[table-format=-1.3(4)]
				S[table-format=-1.3(4)]
				S[table-format=-2.2(3)]
			}
			\toprule
			{PTL}   
				& {$\overline{\delta D} / \SI{}{\micro\meter} (^{+18}_{+08}) $}
				& {$\overline{\delta x} / \SI{}{\micro\meter} (\pm 15)$}
				& {$\overline{\delta y} / \SI{}{\micro\meter} (\pm 15)$}
				& {$\overline{\delta z} / \SI{}{\micro\meter} (\pm 15)$}
				& {$\overline{\delta\theta} / \degree (\pm 0.03)$}
				& {$\overline{\delta\varphi} / \degree (\pm 0.03)$}\\
			\midrule
			00  
				& 10.7 (18)
					& 0.4 (116)
						& -3.8 (72)
							& 0.2 (111)
								& -0.002 (25)
									& -0.339 (384) \\
			01
				& 11.4 (13)
					& -9.3 (66)
						& -0.7 (137)
							& 0.3 (083)
								& -0.033 (13)
									& 0.167 (321) \\
			02
				& 11.2 (31)
					& -7.3 (155)
						& -0.2 (120)
							& 0.7 (117)
								& -0.028 (35)
									& 0.029 (478) \\
			03
				& 3.2 (9)
					& -4.8 (60)
						& -3.6 (123)
							& 0.4(88)
								& -0.021 (12)
									& -0.161 (236) \\
			04
				& 6.2 (24)
					& -11.0 (100)
						& -1.9 (181)
							& 0.9 (105)
								& -0.043 (25)
									& 0.056 (714) \\
			05
				& 6.7 (7)
					& -2.7 (119)
						& -4.0 (85)
							& 0.2 (103)
								& -0.014 (11)
									& -0.230 (343) \\
			06
				& 8.3 (21)
					& -2.2 (107)
						& -2.3 (86)
							& 0.2 (108)
								& -0.010 (11)
									& -0.051 (410) \\
			07
				& 12.3 (10)
					& -2.8 (111)
						& -0.8 (99)
							& 0.2 (121)
								& -0.010 (12)
									& -0.054 (379) \\
			08
				& 12.9 (9)
					& -4.3 (49)
						& -3.1 (124)
							& 0.4 (106)
								& -0.018 (13)
									& -0.122 (231) \\
			09
				& 14.4 (7)
					& -2.7 (101)
						& -3.0 (96)
							& 0.2 (101)
								& -0.013 (8)
									& -0.181 (245) \\
			10
				& 12.0 (6)
					& -3.3 (122)
						& -3.7 (74)
							& 0.2 (115)
								& -0.015 (8)
									& -0.237 (216) \\
			11
				& 8.9 (8)
					& -4.3 (99)
						& -4.1 (85)
							& 0.3 (119)
								& -0.020 (12)
									& -0.199 (358) \\
			\bottomrule
		\end{tabular} 
	\end{table}
	
\begin{figure}[phtb]
		\centering
		\includegraphics[width=\textwidth]{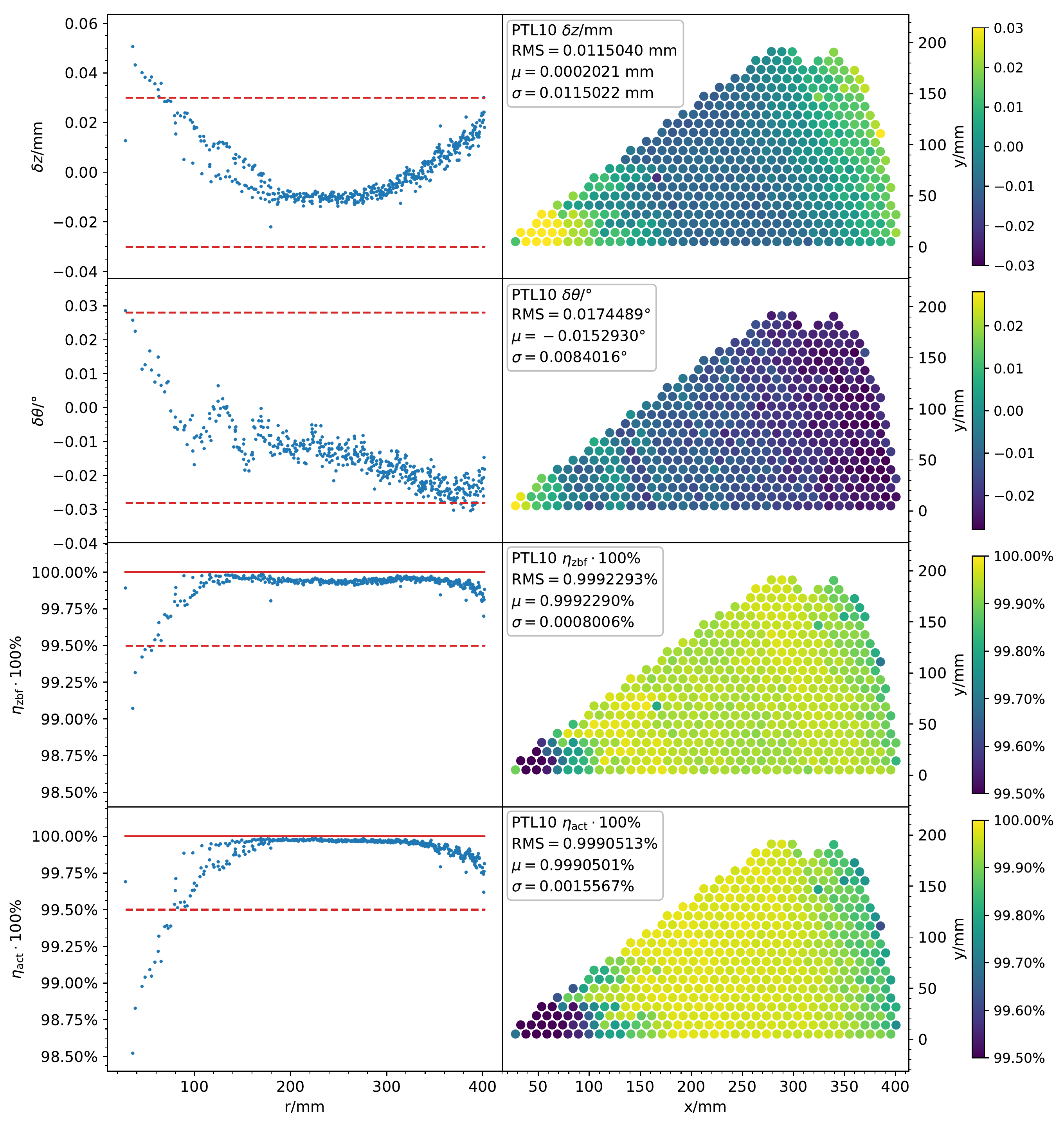}
		\caption{
			\label{fig:petal_metrology_plots}
			(Colour online) Overview of key petal metrology features and resulting throughput after final alignment for PTL10 as an example.
			From top to bottom, the first 2 rows show the spotface centre $z$ deviations ($\delta z$) and nutation angle deviations ($\delta\theta$) in the best-fit (ZBF) alignment; the 3rd row shows the theoretical maximum of optical throughput found in the ZBF alignment ($\eta_\text{zbf}$), and the 4th rows reveals the actual throughput calculated from the final alignment ($\eta_\text{act}$). Left plots are data vs radial distance $r \equiv \sqrt{x^2 + y^2}$; right plots are colour representations of features for 514 individual holes in the physical 2D plane (top view of petal). Root-mean-square, mean, and standard deviation of the data values are annotated as RMS, $\mu$ and $\sigma$. RMS and $\sigma$ are not equal due to nonzero sample mean. Red dashed lines indicate the specified upper and lower tolerances and solid lines are hard limits, as throughput by definition is capped at 100\%.
			}
	\end{figure}

	Petal metrology results of hole features are summarised in Table \ref{table:petal_metrology}, along with mean throughput values from tilt and defocus calculations in the best-fit alignment (ZBF). The top two panels in Fig. \ref{fig:petal_metrology_plots} show more detailed sample plots for two key features, spotface centre $z$ and nutation angle. In the default ABC alignment set by three datums, over half of the petal metrology data which are coordinate system-dependent were out of tolerances. By switching to the ZBF alignment, few were out of tolerances, and the petal quality was assessed in the ZBF alignment. The $3\sigma$ clipping filter for the cylinder fit proved necessary in removing a large volume of outliers, as the \SI{0.3}{\milli\meter} ruby styli were very sensitive to any burr or anodising fragment inside the bores and the soft anodising could easily be scratched by the fine ruby. The anodising had a specified thickness (a.k.a. build-up) of \SI{2.5}{\micro\meter}, meaning \SI{1.25}{\micro\meter} ingress and \SI{1.25}{\micro\meter} growth (net build-up). We found a \SI{2.5}{\micro\meter} reduction in bore diameter on average, which agrees very well with the specification.
	
	In terms of RMS optical throughput, all petals easily meet the science requirement of 99.5\% if mounted ideally with 6 degrees of freedom (the last column in Table \ref{table:throughput}). Although some statistics are slightly out of tolerance on a few petals, they either do not fully represent the petal quality due to CMM metrology systematics, or influence the throughput minimally. A systematic difference was found between diameter measurements by CMM and manual 3-point holtest for the 1400 holes sampled\textemdash the holtest values always larger by about \SI{5}{\micro\meter}. This discrepancy is attributed to the surface roughness of the anodising layer, as the CMM styli touch the surface very gently whereas hand tools, similar to the positioners to be installed into the holes, puts much more pressure on the anodised surface. By the holtest standard, actual bore diameters meet the specifications, and lie on the tight side, which was intended to reduce positioner free play inside the hole. In fact, surface roughness contributes to all CMM measurements made on anodised surfaces, and is not a concern in the vast majority of CMM applications. For the features we analysed, only spotface $z$ is affected (higher than actual by a negligible offset of about \SI{2}{\micro\meter}), and not much for $x$, $y$ or the angles, because the centre of concentric circles remains the same, so does the cylinder axis of coaxial cylinders. This issue does not significantly shift our quantitative results, nor does it change our qualitative conclusions. Relatively large deviations in precession angle were entirely expected, as $\varphi$ depends on the projection of the cylinder axis onto the $xy$-plane. For a precision bore of height $h=\SI{20.5}{\milli\meter}$ at a nutation angle $\theta\in[0.25\degree, 5.88\degree]$, the length of the projected axis is $h\sin\theta \in [\SI{0.09}{\milli\meter}, \SI{2.1}{\milli\meter}]$, corresponding to an uncertainty of about $\SI{0.3}{\degree}$ for a CMM accuracy of $\pm \SI{5}{\micro\meter}$. In fact, given the excellent statistics for other features, we believe the precession angles are much closer to the specification than indicated by CMM data. All 12 production petals were deemed excellent in machining quality and accepted as official DESI parts, out of which ten will be integrated with 5000 positioners and get on sky, while the other two will be spare parts.
	
	\begin{table}[tbh]
		\centering
		\caption{
			\label{table:throughput}
			The projected RMS optical throughput $\widetilde{\eta} = (\frac{1}{514} \sum_{i=1}^{514} \eta_i^2)^{1/2}$ for 12 production petals in several alignments, as calculated from inspection data and our optical model. Columns two to four are the hypothetical best-fit alignment (ZBF), 514-spotface-fit alignment (SPT), and throughput-optimised alignment (TPT). The last column shows the actual, as-aligned results.
			}
		\begin{tabular}{
				c
				S[table-format=-2.3(3)]
				S[table-format=-2.3(3)]
				S[table-format=-2.3(3)]
				S[table-format=-2.3(3)]
			}
			\toprule
			{PTL}   & {$\widetilde{\eta}_\text{zbf}/\% (^{+0.0}_{-0.5}) $}
					& {$\widetilde{\eta}_\text{spt}/\% (^{+0.0}_{-0.5}) $}
					& {$\widetilde{\eta}_\text{tpt}/\% (^{+0.0}_{-0.5}) $}
					& {$\widetilde{\eta}_\text{act}/\% (^{+0.0}_{-0.5}) $}\\
			\midrule
			00  & 99.916 (115)  & 99.909 (175)  & 99.922 (123)  & 99.908 (122)\\
			01  & 99.914 (184)  & 99.914 (184)  & 99.917 (185)  & 99.872 (246)\\
			02  & 99.878 (241)  & 99.870 (276)  & 99.882 (238)  & 99.864 (268)\\
			03  & 99.930 (58)   & 99.928 (73)   & 99.931 (70)   & 99.897 (126)\\
			04  & 99.875 (83)   & 99.874 (91)   & 99.876 (83)   & 99.843 (209)\\
			05  & 99.930 (74)   & 99.927 (102)  & 99.935 (90)   & 99.892 (144)\\
			06  & 99.934 (84)   & 99.931 (123)  & 99.936 (97)   & 99.850 (292)\\
			07  & 99.927 (108)  & 99.924 (154)  & 99.926 (124)  & 99.824 (279)\\
			08  & 99.922 (78)   & 99.921 (100)  & 99.923 (94)   & 99.905 (158)\\
			09  & 99.938 (67)   & 99.936 (93)   & 99.940 (87)   & 99.911 (151)\\
			10  & 99.923 (80)   & 99.919 (123)  & 99.924 (110)  & 99.905 (156)\\
			11  & 99.910 (98)   & 99.907 (141)  & 99.913 (117)  & 99.903 (164)\\
			\bottomrule
		\end{tabular} 
	\end{table}
	
	\begin{figure}[phtb]
		\centering
		\includegraphics[width=\textwidth]{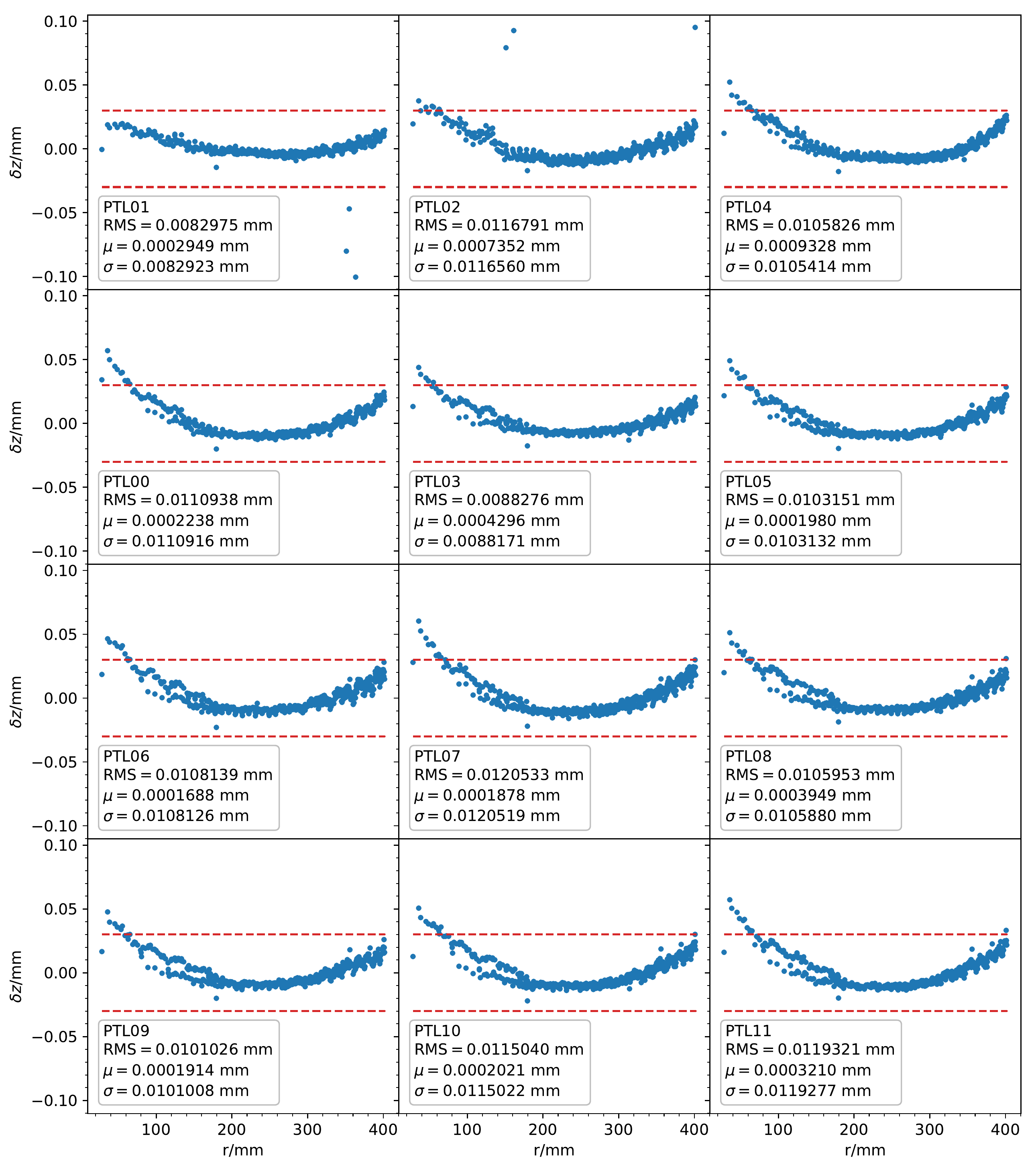}
		\caption{
			\label{fig:actual_z_all_12}
			Spotface $z$ deviations in the ideal (ZBF) alignment for all 12 production petals show excellent machining quality. Subplots are arranged in the chronological sequence the petals were machined, and to be viewed in the row-major order. The 6 sparse outliers for PTL01 and PTL02 were manually confirmed to be measurement errors of the CMM due to the stylus climbing above or falling off the spotface, and are not truncated in the plot for completeness. The resemblance among all 12 plots is remarkable, especially for the last 11 petals, illustrating that machining of aluminium with 5-axis machines is a highly consistent and reproducible process, which has the potential of being further fine-tuned to remove the bowing pattern and attain a flat data trend.}
	\end{figure}
	
	The measured spotface centre $z$ deviations in the ZBF alignment are plotted in their entirety for 6168 holes of 12 official petals in Fig. \ref{fig:actual_z_all_12}, where the surface is concave up relative to the nominal spherical dome in all cases in a highly consistent, reproducible manner. Optical throughput is sensitive to spotface centre $z$ positions and nutation angles: spotface centre $z$ dominates defocus and determines if the fibre tip is put in the depth of focus of the corrector optics, while nutation angle dominates the tilt and determines if the numerical apertures of the corrector optics and of the fibre are aligned. The DESI corrector has a $f/\#$ of 3.86 averaged over the field of view (FOV), and Polymicro FBP fibres have a numerical aperture $\mathrm{NA} = 0.22$, meaning there is no significant throughput loss for small angular tilts due to machining imperfections. On the other hand, the DESI depth of focus for wavelengths $\lambda \in [\SI{360}{\nano\meter}, \SI{980}{\nano\meter}]$ is around \SI{60}{\micro\meter}, the same as the range of errors for spotface centre $z$. Thus spotface centre $z$ error has a larger impact on the throughput compared with nutation angle error, and was the most important feature in petal machining and alignment. The bowing effect in the vertical direction of petals in Fig. \ref{fig:actual_z_all_12} is attributed to the fact that clamping and drilling exerted great stress on aluminium and resulted in a slightly curved petal when the stress was relieved. This in part explains the relatively low throughput near the nose of each petal. Although aluminium machining has its obvious limitations, in light of its remarkable reproducibility as seen in Fig. \ref{fig:actual_z_all_12}, we consider it completely feasible to artificially compensate for highly reproducible and predictable bowing either a priori or through trial-and-error if necessary, such that the machined part turns out flat to \SI{10}{\micro\meter} or even better. In other words, one could intentionally deform the model design and specify carefully calculated ``wrong" numbers to make the part turn out perfect after all stress is relieved. This machining technique should be of particular interest to future surveys which are designed with smaller $f/\#$, shallower depth of focus, and higher precision requirements than DESI.
	
	A comparison of three alignments, ZBF and two other custom alignments, SPT and TPT, is shown in Table \ref{table:throughput}. All three alignments produced very similar throughputs, with ZBF higher than SPT and TPT being the highest. These results perfectly confirmed our expectations; ZBF alignment fits to 514 bores and 514 spotfaces simultaneously whereas SPT only fits to 514 spotface centres and ought to be inferior to ZBF; TPT alignment by definition aims at optimising throughput only with no geometric constraints taken into account. Despite slightly higher throughput, due to the fact that throughput is insensitive to rotations in the $xy$-plane, TPT alignment tends to sacrifice $x$ and $y$ positions to trade for higher throughput, introducing larger than desired $xy$ offsets. The ZBF alignment provides excellent throughput while maintaining geometric fit, and was chosen as the optimal alignment to be pursued, as the differences in throughput between these alignments are nearly negligible. 
	
	Pure aluminium construction of the FPR means it would be soft. Flatness of the FPR datum A was measured as \SI{35}{\micro\meter} when the ring was supported by five jacks and \SI{15}{\micro\meter} when laid directly on the CMM granite. In integration tests, petal mounting was a highly reproducible process with accessories held fixed. The datum tooling ball shifts were within a few microns. With petals installed, the ring may become warped and depending on the load and gauge block, flatness of datum A may be up to \SI{150}{\micro\meter}. In the final alignment with all 10 petal mounting locations occupied, the datum A flatness was \SI{43}{\micro\meter}. It was observed that in addition to uneven load, inappropriate torque on the 8 bolts holding the petal to the ring also significantly warped the FPR and changed petal orientations. A new finite-element analysis was ran and source of distortion identified. The torque specifications were modified to reduce distortion such that the torque of the $45\degree$ bolts is 5 times that of the radial bolts, totalling zero rotational torque on the petal.

	\begin{figure}[tb]
		\centering
		\includegraphics[width=\textwidth]{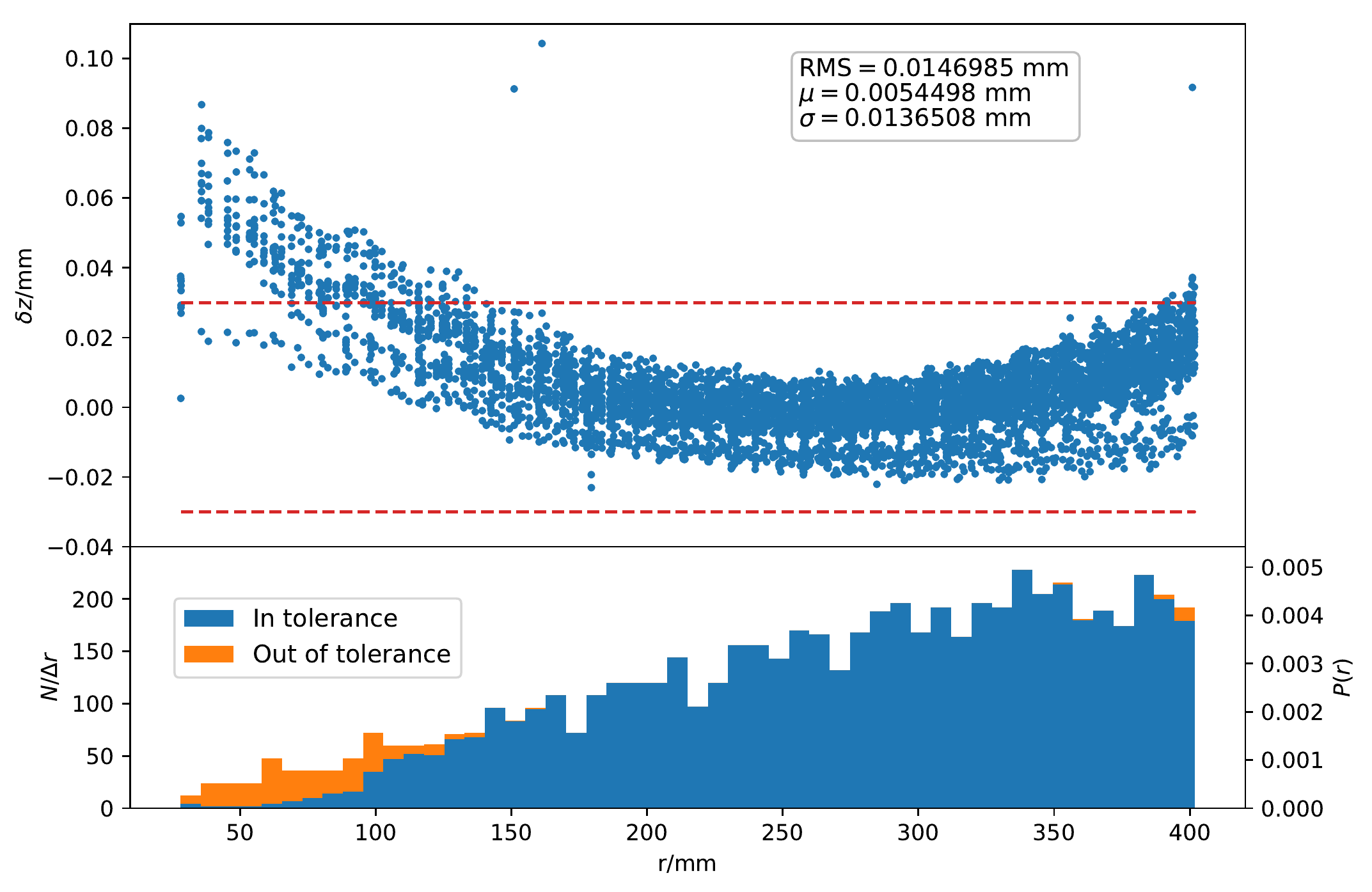}
		\caption{
			\label{fig:actual_z_combined}
			(Colour online) Actual spotface centre $z$ deviation as aligned for 12 production petals in the shared focal plate coordinate system. Data for all 6168 holes are plotted except the 3 outliers of PTL01 due to CMM measurement errors, located below the data cluster outside the plot range and ignored in this plot. The density of data points per unit $r$ from left to right is approximately $\propto 2\pi r \Delta r$, where $\Delta r$ is the constant bin width. This is also evident in the bottom stacked histogram, where the left $y$-axis is the number count per $r$ bin, and the right $y$-axis labels the normalised probability density distribution. Because of the smaller area near $r=0$, although most petals were slightly tilted nose up, only a small number of spotfaces near the centre are outside the $\pm \SI{30}{\micro\meter}$ red dashed window.}
	\end{figure}
	
	The optimiser script was consistently able to find 6 transformation parameters to excellent accuracy when fitting to data and finding the theoretical and actual alignments, with residual sum of squares (RSS) around \SI{0.01}{\milli\meter\squared}. A detailed comparison for 514 holes of PTL10 between the ideal and the actual alignments is included in the last two panels of Fig. \ref{fig:petal_metrology_plots}, where the throughput distributions in the two panels are visually identical. The most important alignment feature, the actual spotface $z$ deviations as aligned for 6168 holes of 12 official petals all combined, is visualised in Fig. \ref{fig:actual_z_combined}. Again, the overall concave-up profile is prominently consistent across 12 official petals. With a constant area-density of holes in the focal plate, the centre of the focal plate has a smaller area and fewer number of holes compared with the outskirts, so despite a small number of holes near the centre being higher than desired in the $z$ coordinate, the vast majority of holes received excellent alignment.
	
\begin{figure}[ptb]
		\centering
		\includegraphics[width=0.83\textwidth]{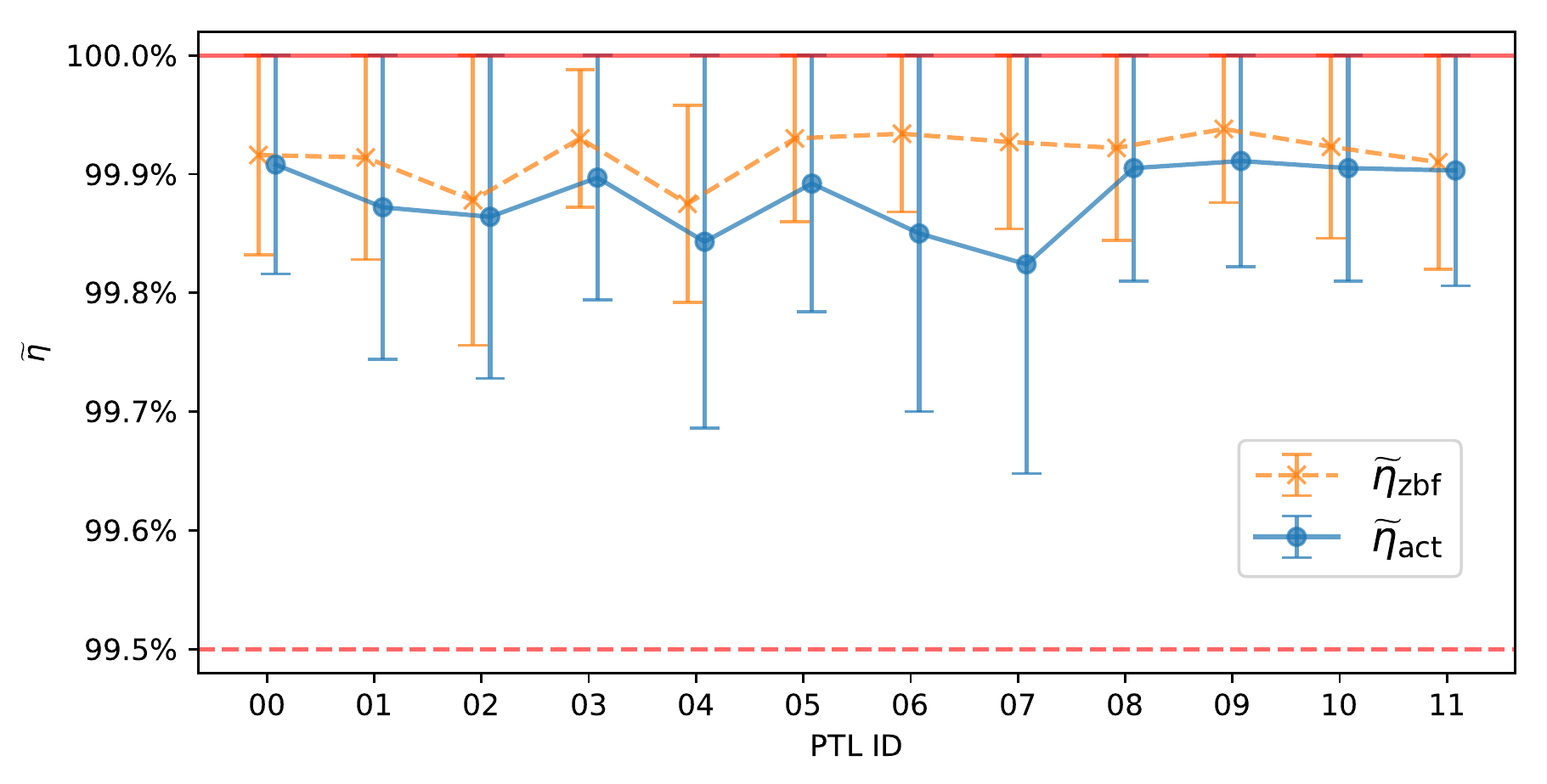}
		\caption{
			\label{fig:throughput_plot}
			RMS throughput of 514 holes in the final alignment for 12 production petals. Blue solid line with circles is the actual RMS throughput, while the orange dashed line with crosses is the ideal RMS throughput in the best-fit (ZBF) alignment representing the best throughput possibly achievable. For each petal, the two data points are slightly offset horizontally only to prevent The error bars from overlapping. The final alignment produced throughput results very close to the upper limits and well above the 99.5\% requirement.
			\vspace{0.3cm}}
	\end{figure}

	Complete final alignment results in terms of throughput are listed in Table \ref{table:throughput} under the actual RMS throughput $\widetilde{\eta}_\text{act}$ column, as well as plotted in Fig. \ref{fig:throughput_plot} along with the best possible throughput from ZBF alignment. Clearly, the as-aligned throughput values of all 12 petals are quite close to the upper bounds. For the spatial dependence of throughput, a coloured 2D view of the throughput of 514 holes plotted in their physical $xy$ coordinates is included in Fig. \ref{fig:petal_metrology_plots}, where the ideal (ZBF) and actual (ACT) throughputs are compared.
	
	GFA-FPP mount plate alignment results from the final run are summarised in Table \ref{table:gfa}. The RMS normal deviations calculated in two ways, $\widetilde{\delta d}_\perp$ and $\widetilde{\delta d ^\prime}_\perp$, were essentially identical up to the 4th decimal place (\SI{0.1}{\micro\meter}), and the difference is negligible compared to CMM accuracy. Naturally there was no strong correlation in the limited sample between any pair of the last three columns, because they track different sets of degree of freedoms. $\delta i$ is the tip out of plane only; $\widetilde{\delta d}_\perp$, as well as $\widetilde{\delta d^\prime}_\perp$, tracks the three d.o.f out of plane: tip (pitch), tilt (roll), and normal translation;  $\widetilde{\delta d}_{rss}$ captures all 6 d.o.f, including the 3 out of plane, and rotation (yaw) and radial translation in plane. Although $\widetilde{\delta d}_{rss}$ approaches \SI{150}{\micro\meter} in some cases, most of it came from the in-plane contribution, and the out-of-plane component at a level of \SI{25}{\micro\meter} or less is sufficiently small. The small inclination angle values also confirmed that the GFA-FPP was in good alignment. 
	
	\begin{table}[tbh]
		\centering
		\caption{
			\label{table:gfa}
			GFA-FPP mount plate alignment results for 11 production petals. Columns are the normal deviations from nominal plane of datums 1, 2, and 3, RMS normal deviations of three datums calculated in two ways, true RMS vector deviations of three datums, and inclination angle deviation. All uncertainties in this table are measurement errors subject to intrinsic CMM accuracy, and no statistical error is computed as only the final best run is shown.}
		\begin{tabular}{
				c
				S[table-format=-2.3, round-mode=places, round-precision=3]
				S[table-format=-2.3, round-mode=places, round-precision=3]
				S[table-format=-2.3, round-mode=places, round-precision=3]
				S[table-format=-2.3, round-mode=places, round-precision=3]
				S[table-format=-2.3, round-mode=places, round-precision=3]
				S[table-format=-2.3, round-mode=places, round-precision=3]
				S[table-format=-2.3, round-mode=places, round-precision=3]
			}
			\toprule
			{GFA-FPP} 
				& {$\delta d^1_\perp / \SI{}{\milli\meter}$}
				& {$\delta d^2_\perp / \SI{}{\milli\meter}$}
				& {$\delta d^3_\perp / \SI{}{\milli\meter}$}
				& {$\widetilde{\delta d}_\perp / \SI{}{\milli\meter}$}
				& {$\widetilde{\delta d ^\prime}_\perp / \SI{}{\milli\meter}$}
				& {$\widetilde{\delta d}_{rss} / \SI{}{\milli\meter}$} 
				& {$\delta i / \degree$} \\
			\midrule
			00  & 0.01587206    & 0.02755246    & 0.02663081    & 0.02394619
				& 0.024         & 0.07027728    & 0.018      \\
			01  & 0.01239256    & 0.0161361     & 0.01906463    & 0.01609771
				& 0.016         & 0.09943441    & 0.004      \\
			02  & 0.00641437    & 0.00285232    & 0.00897081    & 0.00657661
				& 0.007         & 0.10748951    & -0.009     \\
			03  & -0.00410731   & 0.01265299    & 0.00894725    & 0.00925602
				& 0.009         & 0.14335243    & 0.028       \\
			04  & 0.00504961    & 0.01720162    & 0.00876503    & 0.01152129
				& 0.011         & 0.08056594    & 0.021      \\
			05  & 0.00443919    & 0.0080598     & 0.00450135    & 0.00591408
				& 0.006         & 0.08563125    & 0.005      \\
			06  & 0.00769254    & 0.02205404    & 0.00343377    & 0.01363019
				& 0.014         & 0.09138365    & 0.027      \\
			07  & -0.00757413   & 0.00321443    & 0.0064521     & 0.00603682
				& 0.006         & 0.07163707    & 0.016      \\
			08  & 0.01208131    & 0.01157657    & 0.01490586    & 0.01293779
				& 0.013         & 0.08074033    & -0.003     \\
			09  & 0.01179772    & 0.0200852     & 0.02228346    & 0.01861141
				& 0.019         & 0.0614898     & 0.012     \\
			10  & 0.01753585    & 0.03247352    & 0.02110914    & 0.02454677
				& 0.025         & 0.0627012     & 0.026      \\
			\bottomrule
		\end{tabular} 
	\end{table}

\section{Conclusions}
\label{section:conclusions}

	In the era of precision cosmology, instruments such as DESI have to be constructed with micron-level precision. CAD-enabled, fully automated CMM metrology makes it possible to verify the machining quality of extremely complicated precision parts in a highly repeatable and efficient manner, and plays a pivotal role in the integration and alignment. With fixed datums and detailed data down to every single feature, analysis of the transformation and deformation of parts in the optical model directly relates to the optical throughput of the subsystem, and guides the alignment of optomechanical components. We performed complete metrology on the DESI focal plate structure and successfully aligned 12 production petals and 11 GFA-FPP mount plates. Overall, the petals and the ring were machined with excellent quality and the alignment error was $\pm \SI{15}{\micro\meter}$ RMS by geometric metrics and $99.88 \pm 0.12 \%$ RMS throughput by science performance metrics. These positive results demonstrated that our approach to alignment is capable of delivering precision on the \SI{10}{\micro\meter} level and meeting the most demanding alignment challenges for instruments of size on the meter scale.

\section*{Acknowledgements}
	
	The authors thank Eric Buice of LBNL for his contribution to early focal plate integration efforts and useful discussions, and David Yeagley of LBNL for his assistance with CMM metrology. DYT and SPA are supported by the U.S. Department of Energy Office of Science through high-energy physics grant DE-SC0015628. This work was earlier presented at the SPIE Astronomical Telescopes + Instrumentation 2018 conference in Austin, TX\cite{desi_fps_alignment}.
	
	DESI is supported by the Director, Office of Science, Office of High Energy Physics of the U.S. Department of Energy under Contract No. DE–AC02–05CH1123, and by the National Energy Research Scientific Computing centre, a DOE Office of Science User Facility under the same contract; additional support for DESI is provided by the U.S. National Science Foundation, Division of Astronomical Sciences under Contract No. AST-0950945 to the National Optical Astronomy Observatory; the Science and Technologies Facilities Council of the United Kingdom; the Gordon and Betty Moore Foundation; the Heising-Simons Foundation; the National Council of Science and Technology of Mexico, and by the DESI Member Institutions: Aix-Marseille University;  Argonne National Laboratory; Barcelona Regional Participation Group; Brookhaven National Laboratory; Boston University; Carnegie Mellon University; CEA-IRFU, Saclay; China Participation Group; Cornell University; Durham University;  École Polytechnique Fédérale de Lausanne; Eidgenössische Technische Hochschule, Zürich;  Fermi National Accelerator Laboratory;  Granada-Madrid-Tenerife Regional Participation Group; Harvard University; Korea Astronomy and Space Science Institute; Korea Institute for Advanced Study; Institute of Cosmological Sciences, University of Barcelona; Lawrence Berkeley National Laboratory; Laboratoire de Physique Nucléaire et de Hautes Energies; Mexico Regional Participation Group; National Optical Astronomy Observatory; Ohio University; Siena College; SLAC National Accelerator Laboratory;  Southern Methodist University; Swinburne University; The Ohio State University; Universidad de los Andes; University of Arizona; University of California, Berkeley; University of California, Irvine; University of California, Santa Cruz; University College London; University of Michigan at Ann Arbor; University of Pennsylvania; University of Pittsburgh; University of Portsmouth; University of Rochester; University of Queensland; University of Toronto; University of Utah; University of Zurich; UK Regional Participation Group; Yale University. The authors are honored to be permitted to conduct astronomical research on Iolkam Du'ag (Kitt Peak), a mountain with particular significance to the Tohono O'odham Nation. For more information, please visit \url{http://desi.lbl.gov/}.


\bibliography{jatis_2018-references-duan} 
\bibliographystyle{spiebib} 


\section*{Biography}

\vspace{2ex}\noindent\textbf{Duan Yutong} is a high-energy physics Ph.D. Student at Boston University. He received his B.S. in physics with honours from Rhodes College in 2013. He has worked on dynamically contrast-enhanced (DCE) MRI at St. Jude Children's Hospital and interstellar medium simulations at Rhodes. His research interests include astrophysical instrumentation and large-scale structures of the universe. He is a member of SPIE and American Physical Society.

\vspace{2ex}\noindent\textbf{Joseph Silber} started off with a B.A. in studio art from Stanford University and worked as an illustrator, drafter, and art teacher before finding interests in engineering. He received his B.S. from San Francisco City College and M.S. from University of California, Berkeley in mechanical engineering. He has worked on detectors for high-energy particle physics before focusing on the focal plane system for DESI. An engineer by day, he is also a fiction writer by night.

\vspace{2ex}\noindent\textbf{Todd Claybaugh} is a mechanical engineer specializing in precision assemblies and opto-mechanics. He received his B.S. degree from University of California, Berkeley in 2009, and has worked in the semiconductor industry on high-power laser optical systems before joining Berkeley Lab to work on DESI.

\vspace{2ex}\noindent\textbf{Steven Ahlen} is a professor of physics at Boston University. He received his B.S. and M.S. from University of Illinois at Chicago in 1970 and 1971, and his Ph.D. from University of California, Berkeley in 1976. He developed and built magnetic spectrometers for cosmic ray physics above the atmosphere and precision muon tracking detectors for ATLAS, which played an important role in the discovery of the Higgs boson. He holds three US patents in detector technology.

\vspace{2ex}\noindent\textbf{David Brooks} is a Principal Research Fellow in optical instrumentation, Department of Physics and Astronomy at University College London, where he gained his Ph.D. degree in 2001. He is an expert in the manufacture of optics and especially in metal. Over the past 34 years he has constructed many world class instruments and has a long history in developing novel polishing techniques for optical production.

\vspace{2ex}\noindent\textbf{Gregory Tarlé} is a professor of physics and applied physics at the University of Michigan. He received his B.S. degree in physics from Caltech in 1972 and his Ph.D. degree in physics from The University of California, Berkeley in 1978.  He is the author of more than 300 journal papers. His current research interests include dark energy, cosmic ray composition and astrophysical instrumentation. He is a Fellow of the American Physical Society.

\listoffigures
\listoftables

\end{spacing}
\end{document}